\documentclass[aps,twocolumn,prb,preprintnumbers,amsmath,amssymb]{revtex4}

\newcommand{\lsim} 
 {\ \raise.35ex\hbox{$<$}\kern-0.75em\lower.5ex\hbox{$\sim$}\ }
\newcommand{\gsim}
 {\ \raise.35ex\hbox{$>$}\kern-0.75em\lower.5ex\hbox{$\sim$}\ }

\usepackage{graphicx}
\usepackage{dcolumn}
\usepackage{bm}
\usepackage{amsmath}
\usepackage{times}
\usepackage[usenames]{color}
\usepackage{natbib}
\usepackage{ulem}
\allowdisplaybreaks[4]

\begin{document}
\title{Electronic State and Optical Response in a Hydrogen-Bonded Molecular Conductor}
\author{Makoto Naka$^1$ and Sumio Ishihara$^2$}
\affiliation{$^1$Waseda Institute for Advanced Study, Waseda University, Tokyo 169-8050, Japan}
\affiliation{$^2$Department of Physics, Tohoku University, Sendai 980-8578, Japan}
\date{\today}
\begin{abstract} 
Motivated by recent experimental studies of hydrogen-bonded molecular conductors $\kappa$-{\it X}$_3$(Cat-EDT-TTF)$_2$ [{\it X}=H, D], interplays of protons and correlated electrons, and their effects on magnetic, dielectric, and optical properties, are studied theoretically. 
We introduce a model Hamiltonian for $\kappa$-{\it X}$_3$(Cat-EDT-TTF)$_2$, in which molecular dimers are connected by hydrogen bonds. 
Ground-state phase diagram and optical conductivity spectra are examined by using the mean-field approximation and the exact diagonalization method in finite-size cluster. 
Three types of the competing electronic and protonic phases, charge density wave phase, polar charge-ordered phase, and antiferromagnetic dimer-Mott insulating phase are found. 
Observed softening of the inter-dimer excitation due to the electron-proton coupling implies reduction of the effective electron-electron repulsion, i.e. "Hubbard $U$", due to the quantum proton motion. 
Contrastingly, the intra-dimer charge excitation is harden due to the proton-electron coupling.
Implications of the theoretical calculations to the recent experimental results in $\kappa$-{\it X}$_3$(Cat-EDT-TTF)$_2$ are discussed. 
\end{abstract}


\maketitle
\narrowtext



%
%

%




\section{introduction}
It is widely known that the proton is the most lightweight ion and plays essential roles on various physical, chemical and biological phenomena, e.g. quantum ferroelectricity in crystalline solids and liquid crystals, redox reactions in molecules, and self-renewal of DNA in biological materials.~\cite{pauling, schuster, slater, brinc, degennes, prost, lowdin, rein} 
These multifunctional characters of the proton are owing to its high quantum and reactive natures similar to the electrons in solids and molecules. 
Revelation of the entanglements among protons and electrons are widely recognized as one of the central issues in solids, molecules, and biomacromolecules.~\cite{huynh} 
In comparison with the biomaterial systems in which a huge number of components and couplings with them bring about unmanageable complexity, the crystalline solids provide suitable playground for the electron-proton entanglement phenomena. 

\begin{figure}[t]
\begin{center}
\includegraphics[width=1.0\columnwidth, clip]{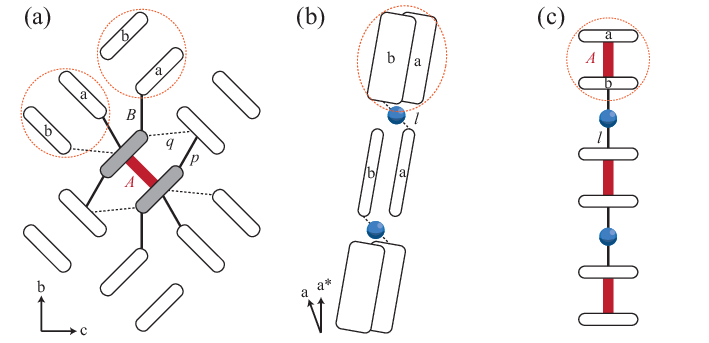}
\end{center}
\caption{
Schematic (a) intra-layer and (b) inter-layer crystal structures of $\kappa$-{\it X}$_3$(Cat)$_2$. 
Ellipses and circles represent the Cat molecules and protons, respectively. 
Solid and dotted lines denoted by $A$, $B$, $p$, $q$, and $l$ are the major inter-molecular bonds with dominant electron transfer integrals and Coulomb interactions. 
The two molecules in the dimers are denoted by $a$ and $b$. 
The $a^*$ axis is perpendicular to the $b$-$c$ plane. 
(c) Lattice structure and the major bonds in the simplified one-dimensional model. 
}
\label{fig:lattice}
\end{figure}

Recently, a new series of organic molecular compounds showing a predominant proton-electron coupling was discovered.~\cite{kamo, isono_nat, isono_prl, ueda, shimozawa} 
The chemical formula of this material is $\kappa$-{\it X}$_3$(Cat-EDT-TTF)$_2$, where Cat-EDT-TTF represents the catecholfused ethylenedithiotetrathiafulvalene (abbreviated as Cat) and {\it X} takes the proton H or deuteron D. 
The crystal structure consists of the Cat molecular layers, shown in Fig.~\ref{fig:lattice}(a), which are connected by the hydrogen bonds with each other. 
The hydrogen-bond network consisted of the dimerized Cat molecules and the protons are extended to the out-of-plane direction as shown in Fig.~\ref{fig:lattice}(b). 
In a Cat molecular dimer, a pair of the molecular orbitals forms the bonding and antibonding orbitals. 
Since one hole exists per the Cat molecular dimer, the antibonding molecular orbital band is identified as a half-filled band. 
Thus, the system is a candidate of a Mott insulator in the case of a strong electron-electron interaction. 
This is termed a dimer-Mott (DM) insulator, which is well known in the low-dimensional organic crystals.~\cite{fukuyama, kino} 
A prototypical example of the DM insulator is the $\kappa$-type bisethylenedithio-tetrathiafulvalene (abbreviated as ET) organic salts.~\cite{kanoda, miyagawa} 
Instead of the anion molecules in the ET salts, the hydrogen bonds connect the conducting layers in the present compounds. 
It is expected that the proton motions affect significantly the electronic states in the Cat layer.  
Thus, the present compounds are recognized as a possible proton-electron entangled system, where not only new electronic and protonic states but also novel functionalities based on their coupling are expected. 

One of the intriguing phenomena in $\kappa$-{\it X}$_3$(Cat)$_2$ with {\it X}$=$H is that it does not show any magnetic long-range ordered states down to $50$ mK in spite that the magnitude of the spin exchange interaction is estimated to be around $80$ K.~\cite{isono_prl} 
This indicates a realization of a quantum spin liquid state.~\cite{shimizu, itou} 
The crystal structural analysis by the X-ray diffraction measurements reveals that the average position of the proton is at the center of the hydrogen bond, suggesting the quantum tunneling of the protons among the two potential minima.~\cite{ueda} 
The electronic and protonic states are drastically changed by deuteration; $\kappa$-D$_3$(Cat)$_2$ undergoes a first order phase transition at around $185$ K from a paramagnetic to non-magnetic insulator. 
This is accompanied by an alternate charge disproportionation between the crystallographically nonequivalent Cat dimers and displacements of the deuterons from the centers of the hydrogen bond.~\cite{ueda} 
These results indicate that the proton degree of freedom in the hydrogen bond strongly interacts with the charge and spin degrees of freedom in the electron in the Cat dimers. 
However, the microscopic picture of the electronic and protonic states in $\kappa$-H$_3$(Cat)$_2$ and $\kappa$-D$_3$(Cat)$_2$ and a role of proton-electron coupling behind it remain to be clarified. 

In this paper, motivated by the recent researches in $\kappa$-{\it X}$_3$(Cat)$_2$, we present a microscopic theory for the hydrogen-bonded molecular conductor, which involves the proton-electron coupling, quantum proton fluctuation, and electron correlation effect in the $\pi$-electron system. 
We introduce an effective model for $\kappa$-{\it X}$_3$(Cat)$_2$, and calculate the ground-state phase diagram and optical spectra by complimentary use of the mean-field approximation and the exact diagonalization method. 
Competition between the proton-electron coupling and electron correlations causes cooperative disproportionation of the electronic and protonic charges. 
In the ground state, a CDW state with the inter-dimer charge disproportionation, polar charge order (CO) states with the intra-dimer charge disproportionation, and an antiferromagnetic (AFM) DM insulating state compete with each other. 
The CDW fluctuation strongly suppresses the AFM spin correlations in the DM phase. 
The charge excitation energies in the DM state are changed significantly when the proton-electron coupling turns on; the intra- and inter-dimer charge excitations show hardening and softening, respectively. 
This softening of the inter-dimer excitation implies a reduction of the effective electron-electron repulsion, i.e. "Hubbard $U$", due to the quantum proton motion.  
On the other hand, the hardening of the intra-dimer excitation is caused by cooperative displacements of protons and electrons. 
These results give us a method to evaluate  the magnitude of the proton-electron coupling through the optical measurements.
The present results provide a fundamental understanding of proton-electron cooperative phenomena in crystalline solids and give a guiding principle to explore new functional hydrogen-bonded molecular conductors. 

This paper is organized as follows. 
In Sec. II, an effective model Hamiltonian for $\kappa$-{\it X}$_3$(Cat)$_2$ is introduced. 
In Sec. III, numerical methods and physical quantities calculated in the following sections are introduced. 
In Sec. IV, the ground-state properties obtained by the mean-field approximation and the exact diagonalization method are presented. 
In Sec. V, analyses of the charge excitations and optical responses are presented. 
Section VI is devoted for discussion and summary. 
%
%
%
\begin{figure}[t]
\begin{center}
\includegraphics[width=1.0\columnwidth, clip]{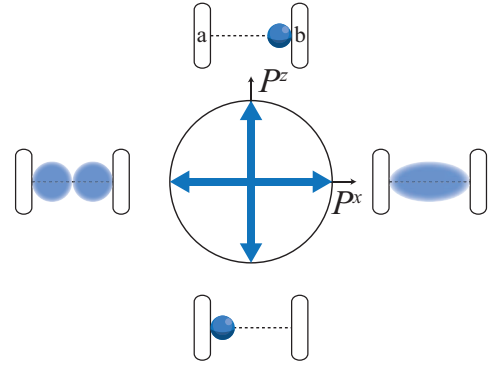}
\end{center}
\caption{
Pseudo-spin directions in the $P^{x}$-$P^{z}$ plane and proton states. 
Shades represent proton distribution in a hydrogen bond, schematically. 
The molecules $a$ and $b$ belong to neighboring dimers connected by the hydrogen bond. 
}
\label{fig:ps}
\end{figure}
%
%
%
\begin{figure}[t]
\begin{center}
\includegraphics[width=1.0\columnwidth, clip]{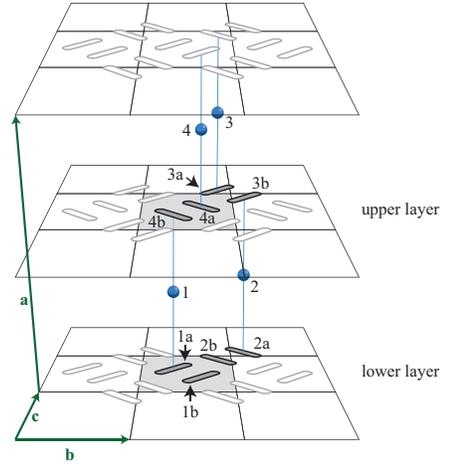}
\end{center}
\caption{ 
Schematic picture of the three-dimensional dimer-proton network and the primitive translational vectors in $\kappa$-{\it X}$_3$(Cat)$_2$. 
Ellipses and circles represent the Cat molecules and protons, respectively. 
The shaded four dimers and four protons denoted by $1$-$4$ are included in the unit cell adopted in the mean-field calculation. 
}
\label{fig:cell}
\end{figure}
%
%
%
\section{Model}
We introduce a tight-binding model for $\kappa$-{\it X}$_3$(Cat)$_2$ where the electron and proton degrees of freedom are taken into account. 
The model Hamiltonian is give by 
\begin{align}
{\cal H} = {\cal H}_{\rm e} + {\cal H}_{\rm pro}, 
\label{eq:hamil}
\end{align}
where the first term represents the kinetic energy and Coulomb interactions in the $\pi$-electron system in the Cat layers and the second term represents the proton-electron coupling and quantum proton motion. 
The first term is given by the extended Hubbard model, 
in which the highest occupied molecular orbitals of the two molecules in the Cat dimer are introduced as the basis orbitals. 
This is given by 
\begin{align}
{\cal H}_{\rm e} 
&= t_{A} \sum_{i \sigma}(c^{\dagger}_{i a \sigma}c_{i b \sigma} + {\rm H. c.})
+ \sum_{\langle ij \rangle \sigma} t^{\mu\mu'}_{ij}(c^{\dagger}_{i \mu \sigma}c_{j \mu' \sigma} + {\rm H. c.}) \notag \\
&+ U\sum_{i \mu}n_{i \mu \uparrow}n_{i \mu \downarrow}
+ V_{A} \sum_{i}n_{i a}n_{i b} + \sum_{\langle ij \rangle \mu \mu'} V^{\mu \mu'}_{ij} n_{i \mu}n_{j \mu'}, 
\label{eq:hamil_e}
\end{align}
where $c_{i\mu\sigma}$ is the annihilation operator of a hole with spin $\sigma=(\uparrow, \downarrow)$ at $\mu(=a, b)$ molecule of the $i$-th dimer, $n_{i \mu \sigma}=c_{i \mu \sigma}^\dagger c_{i \mu \sigma}$ is the number operator, and $n_{i\mu} = \sum_{\sigma} n_{i\mu\sigma}$. 
The first and second terms in Eq.~(\ref{eq:hamil_e}) represent the intra- and inter-dimer electron transfers, respectively. 
The third and fourth terms represent the Coulomb interaction between two holes in the same molecule and that between two holes in the different molecules in the dimer, respectively. 
The last term describes the Coulomb interaction between holes in the different dimers. 

We introduce the pseudo-spin (PS) operator $\bm P$ with the amplitude $1/2$ to describe the proton degree of freedom in the hydrogen bond. 
The eigenstates of $P^{z}$, denoted by $\left| + \right\rangle$ and $\left| - \right\rangle$, represent the states where the proton is located at one side of the two potential minima 
The eigenstates of $P^{x}$, denoted by 
$\left| +^x \right\rangle = (\left| + \right\rangle + \left| - \right\rangle)/\sqrt{2}$ and $\left| -^x \right\rangle = (\left| + \right\rangle - \left| - \right\rangle)/\sqrt{2}$, 
represent the states where the proton occupies the bonding and antibonding states in the double-well potential, respectively. 
Using the PS operators, the second term in Eq.~(\ref{eq:hamil}) is given by 
\begin{align}
{\cal H}_{\rm pro} = 2 t_{\rm pro}\sum_{i} P_i^{x} + \frac{1}{2}g \sum_{\langle ij \rangle} (n_{j a} - n_{i b})P_{i}^{z} , 
\end{align}
where the first term represents the quantum proton tunneling between the two potential minima, and the second term represents the proton-hole coupling. 
The coupling constant is chosen to be repulsive ($g > 0$) because both the hole and proton have the positive charge. 

\section{Method}
We analyze the effective model in Eq.~(\ref{eq:hamil}) by the complimentary two methods, the mean-field approximation and the exact diagonalization method based on the Lanczos algorithm. 
In the mean-field method, we consider the three-dimensional lattice structure of $\kappa$-{\it X}$_3$(Cat)$_2$ shown in Fig.~\ref{fig:cell}. 
In the Lanczos method, we employ a model in the one-dimensional chain which is formed by the Cat dimers connected by the hydrogen bonds shown in Fig.~\ref{fig:lattice}(c). 
This is a minimal structural unit to analyze the proton-electron coupling in $\kappa$-{\it X}$_3$(Cat)$_2$, . 
The details of the methods are explained below.

\subsection{Mean-field approximation}
We examine the charge, spin, and proton configurations in the ground state of the effective Hamiltonian by use of the mean-field approximation. 
The electron-electron interactions and proton-electron coupling in the Hamiltonian given in Eq.~(\ref{eq:hamil}) are decoupled as
\begin{align}
n_{i \mu \uparrow} n_{i \mu \downarrow} \rightarrow &\langle n_{i \mu \uparrow} \rangle n_{i \mu \downarrow} + n_{i \mu \uparrow} \langle n_{i \mu \downarrow} \rangle - \langle n_{i \mu \uparrow} \rangle \langle n_{i \mu \downarrow} \rangle, \notag \\
n_{i \mu \sigma} n_{j \mu' \sigma'} \rightarrow &\langle n_{i \mu \sigma} \rangle n_{j \mu' \sigma'} + n_{j \mu \sigma} \langle n_{j \mu' \sigma'} \rangle - \langle n_{i \mu \sigma} \rangle \langle n_{j \mu' \sigma'} \rangle \notag \\
- \delta_{\sigma \sigma'} ( &\langle c_{j \mu' \sigma'}^\dagger c_{i \mu \sigma} \rangle c_{i \mu \sigma}^\dagger c_{j \mu' \sigma} + c_{j \mu' \sigma}^\dagger c_{i \mu \sigma} \langle c_{i \mu \sigma}^\dagger c_{j \mu' \sigma} \rangle \notag \\
- &\langle c_{j \mu' \sigma}^\dagger c_{i \mu \sigma} \rangle \langle c_{i \mu \sigma}^\dagger c_{j \mu' \sigma} \rangle ), \notag \\
n_{i \mu} P_i^z \rightarrow &\langle n_{i \mu} \rangle P_i^z + n_{i \mu} \langle P_i^z \rangle - \langle n_{i \mu} \rangle \langle P_i^z \rangle.  
\end{align}
The mean-field approximation introduced above is suitable to examine possible ordered states in the system with the multiple degrees of freedom. 
In the numerical calculations, we adopt a unitcell including the four Cat dimers and four protons located between the two Cat layers shown in Fig.~\ref{fig:cell} and $20 \times 20 \times 20$ $\bm k$-points in the Brillouin zone. 

We set the value of the intra-dimer electron transfer $t_{A}=1$ as the unit of energy. 
Through the numerical calculations with several sets of parameter values, we find that the quantum proton tunneling $t_{\rm pro}$, the proton-electron coupling $g$, and the intra-molecular Coulomb interaction $U$ are most relevant parameters to the electron and proton states. 
Therefore, we will show the results by varying these parameter values while fixing the others. 
The values of the dominant electron transfers shown in Figs.~\ref{fig:lattice}(a) and \ref{fig:lattice}(b) are chosen based on the recent first-principles band calculation as $t_{p} = 0.2$, $t_{q} = 0$, $t_{B} = 0.3$, and $t_{l} = 0.1$,~\cite{tsumuraya} and the inter-molecular Coulomb interactions are fixed as $V_{A} = 0.75$ and $V_{p} = V_{q} = V_{B} = 0.5$ assuming $1/r$-type distance dependence.
\subsection{Exact diagonalization}
We apply the Lanczos exact diagonalization method to the Hamiltonian in the one-dimensional lattice shown in Fig.~\ref{fig:lattice}(c) in order to examine the charge and spin correlations and the charge dynamics in the ground state.  
We introduce two kinds of the charge correlation functions characterizing the CDW and the polar CO states defined as
\begin{align}
N(k) &= \frac{1}{N^2}\sum_{ij}\langle n_{i} n_{j} \rangle e^{-i k (r_{i} - r_{j})}, \\
P(k) &= \frac{1}{N^2}\sum_{ij}\langle p_{i} p_{j} \rangle e^{-i k (r_{i} - r_{j})}, 
\end{align}
respectively, and $N$ is the number of the molecular dimers. 
Here, $n_{i} = (n_{ia} + n_{ib} - 1)/2$ and $p_{i} = (n_{ia} - n_{ib})/2$ represent the charge density and electric dipole moment in the $i$-th dimer, respectively, and $r_{i}$ denotes the center of the $i$-th dimer. 
The spin correlation function characterizing the AFM structure is given by 
\begin{align}
S(k) = \frac{1}{N^2}\sum_{ij}\langle s^z_{i} s^z_{j} \rangle e^{-i k (r_{i} - r_{j})},  
\end{align}
where $s^{z}_{i} = (n_{i\uparrow} - n_{i\downarrow})/2$. 
The maximum values of the correlation functions introduced above are $0.25$, when the classical ordered states are realized. 
Charge dynamics is examined by calculating the optical conductivity spectra defined by 
\begin{align}
\sigma(\omega) &= - \frac{e^2}{\omega N} \sum_{m(\neq 0)}  
{\rm Im} \Biggl[ \frac{\left| \left\langle 0 \right| j \left| m \right\rangle \right|^2}{\omega - E_m + E_0 + i \eta} \notag \\ 
&- \frac{\left| \left\langle 0 \right| j \left| m \right\rangle \right|^2}{\omega + E_m - E_0 + i \eta} \Biggr], 
\label{eq:opt}
\end{align}
where $E_{0}$ and $\left| 0 \right\rangle$ represent the ground-state energy and wave function, respectively, $E_{m}$ and $\left| m \right\rangle$ represent the $m$-th eigen energy wave function, respectively, $\eta$ is an artificial broadening factor, and $j = i \sum_{\langle ij \rangle \sigma} t_{ij} (r_{i} - r_{j}) (c_{i\sigma}^{\dagger}c_{j\sigma} - c_{j\sigma}^{\dagger}c_{i\sigma})$ is the electronic current operator. 
The light is applied along the one-dimensional chain. 
We set $t_A=1$ and $V_A=0.75$ and change the parameter values of $g$, $t_{\rm pro}$, and $U$.  
The inter-dimer electron transfer and Coulomb interaction on the bond $l$ shown in Fig.~\ref{fig:lattice}(c) are chosen as $t_l=0.3$ and $V_l=0.5$. 
The finite-size cluster including six dimers and six protons is adopted. 
%
%
%
\begin{figure*}[t]
\begin{center}
\includegraphics[width=2.0\columnwidth, clip]{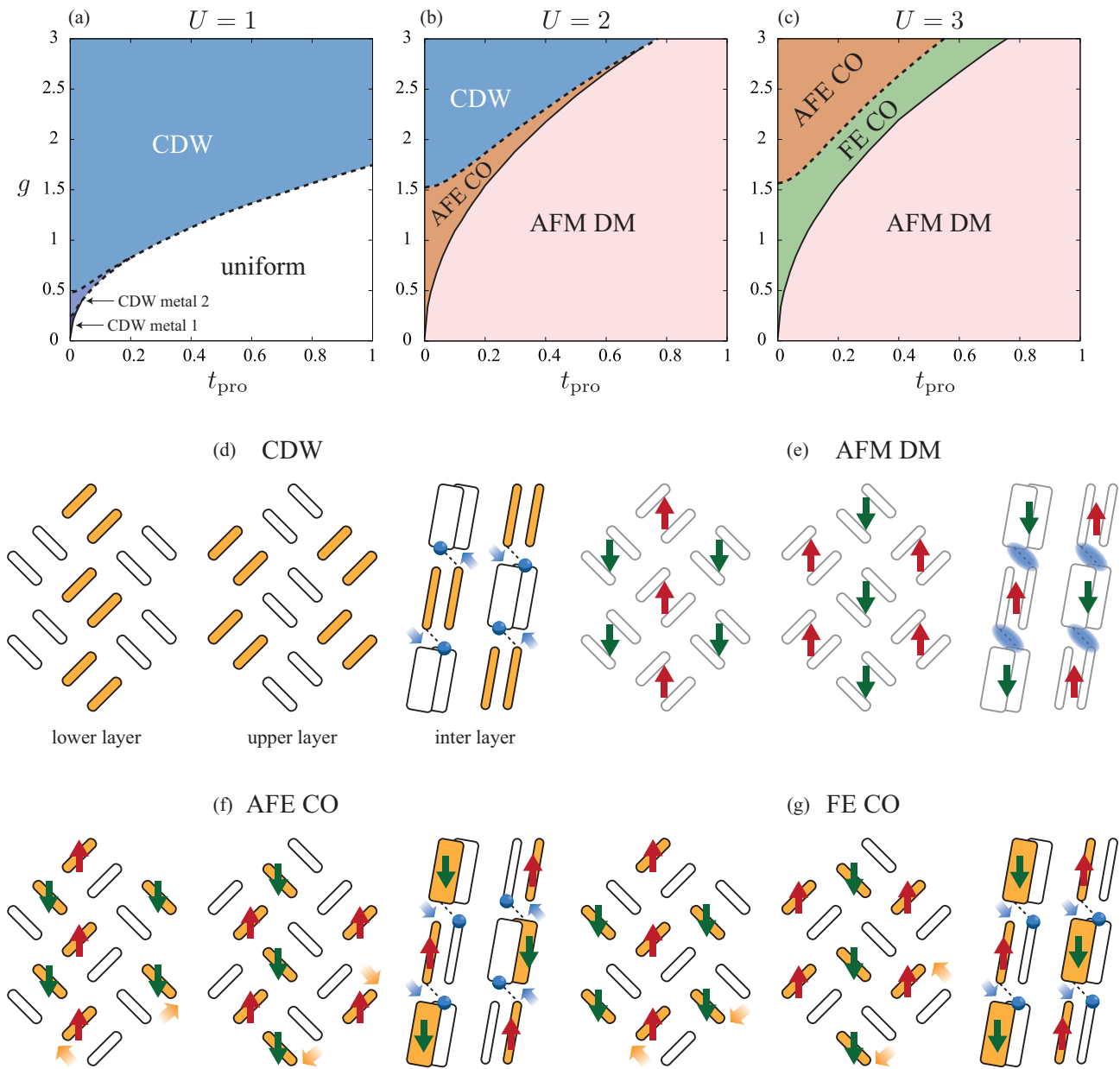}
\end{center}
\caption{
Ground-state phase diagrams obtained by the mean-field approximation at (a) $U=1$, (b) $U=2$, and (c) $U=3$. 
Solid and broken lines denote the second and first order transitions, respectively. 
Schematic charge, spin, and proton configurations in (d) CDW, (e) AFM DM, (f) AFE CO, and (g) FE CO. 
Filled and open ellipses represent the hole-rich and hole-poor molecules, respectively. 
Circles and shaded ellipses between the dimers represent the protons. 
Solid and shaded arrows represent the spin moments and the electric dipole moments due to electrons and protons, respectively.
}
\label{fig:pd}
\end{figure*}
%
%
%
\begin{figure*}[t]
\begin{center}
\includegraphics[width=2.0\columnwidth, clip]{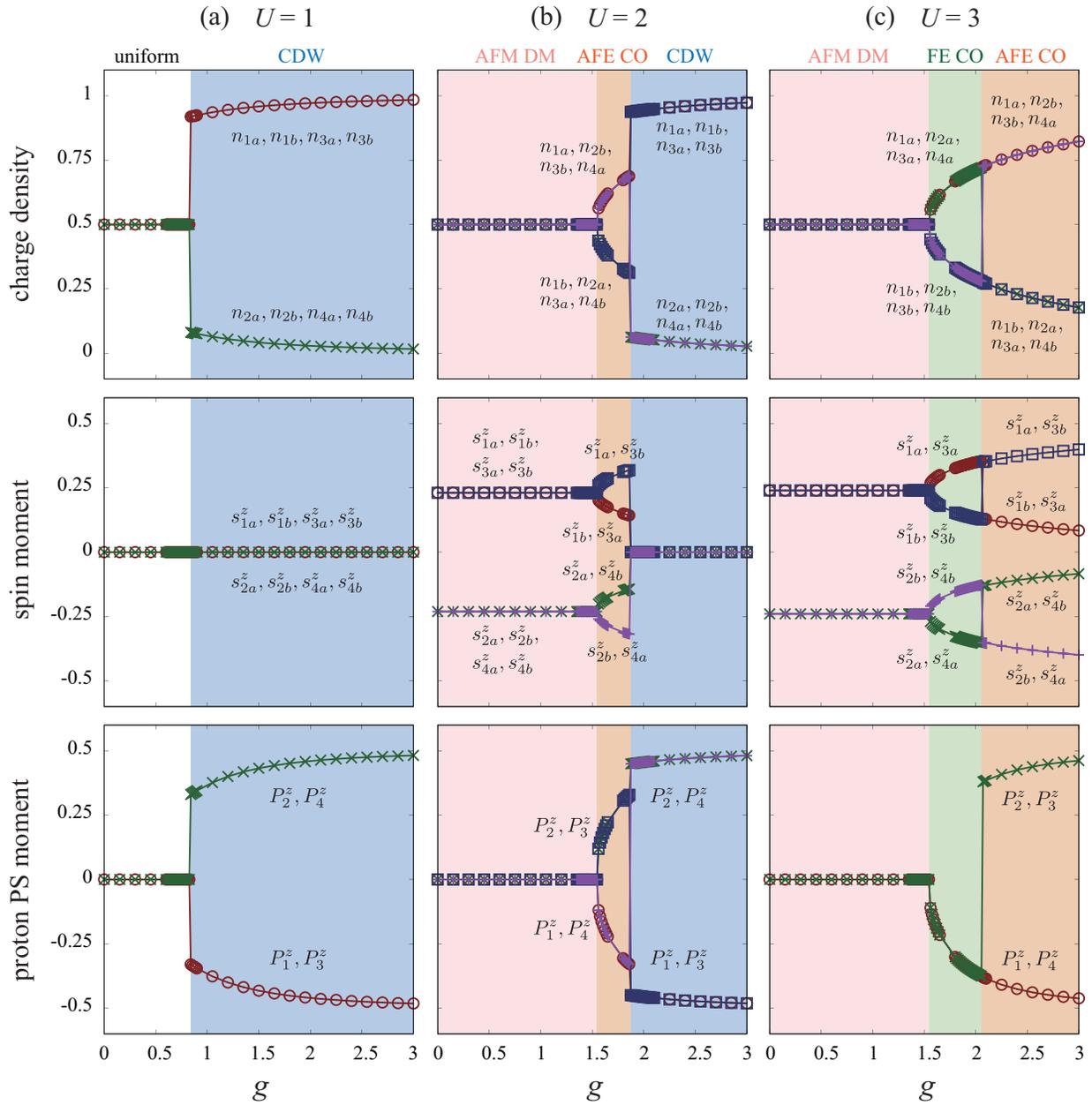}
\end{center}
\caption{
Charge densities (upper panels), spin moments (middle panels), and proton PS (lower panels) at each molecule as functions of $g$ at (a) $U=1$, (b) $U=2$, and (c) $U=3$. 
Parameter values are chosen to be $t_{A}=1$, $t_{B}=0.3$, $t_{p}=0.2$, $t_{q}=0$, $t_{l}=0.1$, $V_{A}=0.75$, $V_{B}=V_{p}=V_{q}=0.5$, and $t_{\rm pro}=0.2$. 
}
\label{fig:order}
\end{figure*}
%
%
%
%
%
%
\section{Ground state}
\begin{figure*}[t]
\begin{center}
\includegraphics[width=2.0\columnwidth, clip]{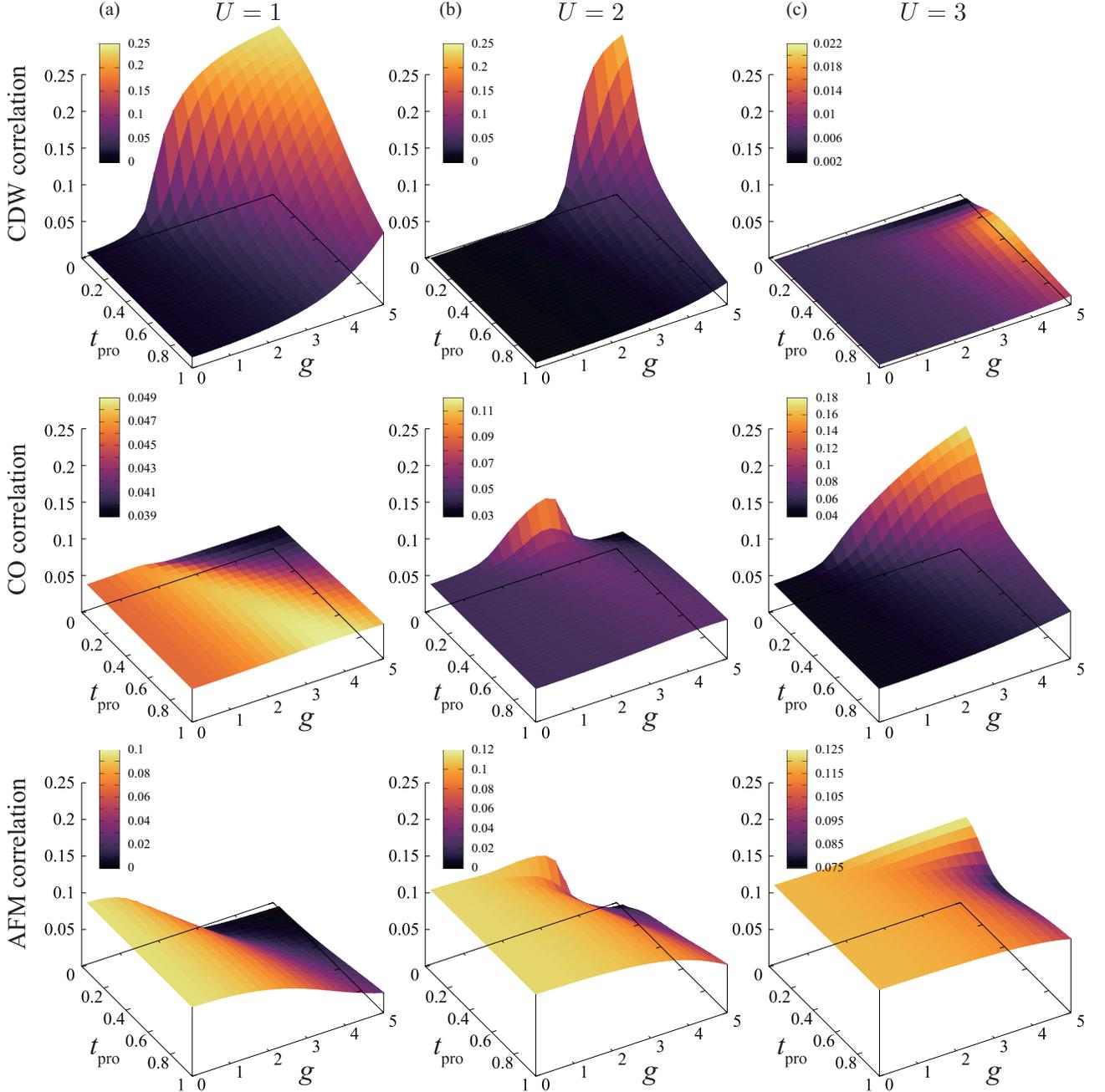}
\end{center}
\caption{
Correlation functions for CDW (upper panels), polar CO (middle panels), and AFM (lower panels) in the ground state of the one-dimensional model at (a) $U=1$, (b) $U=2$, and (c) $U=3$. 
}
\label{fig:corr}
\end{figure*}
%
%
%
\begin{figure}[t]
\begin{center}
\includegraphics[width=1.0\columnwidth, clip]{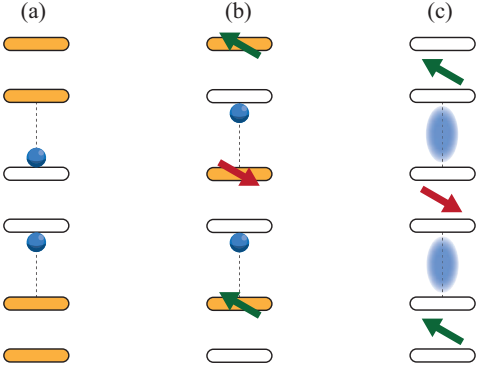}
\end{center}
\caption{
Schematic charge, spin, and proton configurations in (a) CDW, (b) polar CO, and (c) AFM DM states in the one-dimensional model. 
}
\label{fig:1d}
\end{figure}
%
%
%
\begin{figure}[t]
\begin{center}
\includegraphics[width=1.0\columnwidth, clip]{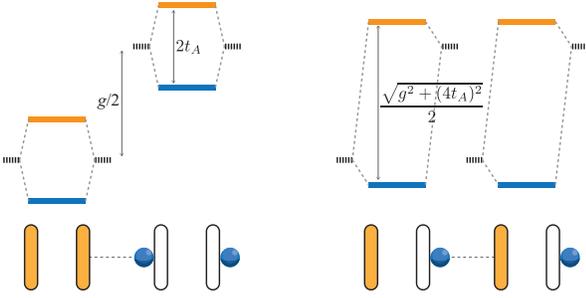}
\end{center}
\caption{
Molecular-orbital energy diagrams in the CDW state (left) and the polar CO state (right). 
Bold solid and broken lines represent the energy levels of the molecular orbitals with and without $t_A$, respectively. 
Lower panels show schematic electronic charge and proton configurations. 
}
\label{fig:level}
\end{figure}
%
%
%
\subsection{Phase diagram}
In this section, we present the ground-state properties obtained by the mean-field approximation. 
Figures~\ref{fig:pd}(a)-\ref{fig:pd}(c) show the ground-state phase diagrams on the plane of the proton-tunneling $t_{\rm pro}$ and the proton-electron coupling $g$. 
The intra-molecular Coulomb interaction is chosen as $U = 1$, $2$, and $3$ in Figs~\ref{fig:pd}(a), \ref{fig:pd}(a), and \ref{fig:pd}(c), respectively. 
We find four kinds of insulating phases with different charge, spin, and proton configurations, termed CDW, AFE CO, AFM DM, and FE CO. 
We also find a uniform metallic phase, where charge and spin densities are uniform, and two metallic CDW phases, termed CDW metal 1 and CDW metal 2. 
Schematic charge and spin configurations in the insulating phases are presented in Figs.~\ref{fig:pd}(d)-\ref{fig:pd}(g). 

Characteristics in each phase is summarized. 
In the CDW phase shown in Fig.~\ref{fig:pd}(d), one of the two kinds of the dimers is occupied by two holes while the other dimer is nearly empty, and spin polarization does not appear in either dimers.  
The protons shift from the center of the hydrogen bonds toward the dimers which are poorly occupied by holes. 
The electric dipole moments in the hydrogen bonds are ordered alternately. 
Thus, the CDW phase is identified as an antiferroelectric (AFE) phase. 
In the AFM DM phase, the charge density in each molecule is 0.5, and spins are antiferromagnetically ordered as shown in Fig.~\ref{fig:pd}(e). 
The protons tunnel between the two potential minima, and their average positions are the center of the hydrogen bonds. 
The other two insulating phases are polar CO phases, where a charge disproportionation occurs in the two molecules inside the dimer, generating an intra-dimer electric dipole moment.~\cite{jawad, lang, drichko, naka, hotta} 
The AFE CO and FE CO are distinguished with each other by the configurations of the intra-dimer electric dipoles and the proton displacements. 
In the AFE CO phase shown in Fig.~\ref{fig:pd}(f), both the two kinds of the electric polarizations owing to the electron and proton orders are zero. 
This phase is identified as an AFE phase. 
On the other hand, in the FE CO phase shown in Fig.~\ref{fig:pd}(g), the electric dipole moments due the electron and proton orders are not canceled out and the macroscopic electric polarization remains in the $a$-$c$ plane, showing a ferroelectric order. 
In both the two CO phases, 
the magnetic moments in the dimers are almost located in one of the molecules in the dimers and are ordered antiferromagnetically. 
In the phase diagrams shown in Figs.~\ref{fig:pd}(a)-\ref{fig:pd}(c), the CDW, AFE CO, and FE CO phases, in which the electronic charge disproportionation in the dimer and the proton displacement are finite, appear in the large-$g$ and small-$t_{\rm pro}$ region. 
On the other hand, the AFM DM and uniform metallic phases, in which the charge distributions are uniform and the protons are not polarized in the hydrogen bonds, are stabilized in the small-$g$ and large-$t_{\rm pro}$ region. 
In addition, two metallic CDW phases appear between the CDW and uniform metallic phases. 
The electronic and protonic charge configurations of the both two phases are similar to those of the CDW phase. 
The CDW metal 1 phase is nonmagnetic, while the CDW metal 2 phase involves a weak staggered magnetic order in between the charge rich and poor dimers. 

To elucidate the variations of the charge, spin, and proton configurations in more detail, we show in Fig.~\ref{fig:order} the $g$ dependences of the charge densities $\langle n_{i \mu} \rangle = (\langle n_{i \mu\uparrow} \rangle + \langle n_{i \mu\downarrow} \rangle)/2$, spin moments $\langle s_{i \mu}^z \rangle = (\langle n_{i \mu\uparrow} \rangle - \langle n_{i \mu\downarrow} \rangle)/2$, and proton PS moments $\langle P_{i}^z \rangle$, where the subscripts $i (=1,2,3,4)$ and $\mu(=a,b)$ denote the sublattices defined in Fig.~\ref{fig:cell}. 
Figure~\ref{fig:order}(a) shows the results at $t_{\rm pro}=0.2$ and $U=1$. 
In the small-$g$ region, the holes are uniformly distributed in each molecule, and neither the spin polarization nor the the proton displacement appear. 
This is interpreted as a uniform metallic phase. 
With increasing $g$, both the charge densities and proton PS moments jump at around $g=0.7$, and show the staggered ordered states in the CDW phase. 

Figure~\ref{fig:order}(b) shows the $g$ dependences of the order parameters at $t_{\rm pro} = 0.2$ and $U=2$. 
As shown in the phase diagram in Fig.~\ref{fig:pd}(b), the uniform metallic phase seen in $U=1$ is replaced by the AFM DM phase. 
The AFE CO phase with the intra-dimer electric dipole emerges between the CDW and AFM DM phases. 
With increasing $g$ from the AFM DM phase, the system changes into the AFE CO phase, where the charge densities and the spin moments show the different values between the $a$ and $b$ molecules inside the dimer unit, and the protons are polarized in the hydrogen bonds. 
This phase transition at around $g = 1.5$ is of the second order. 
With further increasing $g$, the first-order phase transition occurs from the AFE CO phase to the CDW phase. 
At the transition point, the charge densities in the two kinds of dimers and the proton configurations in the one-dimensional dimer-proton chain change from the uniform [Fig.~\ref{fig:pd}(f)] to staggered [Fig.~\ref{fig:pd}(d)] alignments, and the spin moments disappear. 
In the large-$t_{\rm pro}$ region in Fig.~\ref{fig:pd}(b), the AFE CO phase disappears and the phase boundary with the first-order transition between the AFM DM phase to the CDW phase emerges. 

In the phase diagram at $U=3$, the CDW phase shown in $U=2$ is replaced by the AFE CO in the region of large $g$, and FE CO phase appears between the AFE CO and AFM DM phases. 
The phase transition between the AFM DM and FE CO is of the second order as shown in Fig. \ref{fig:order}(c). 
In the FE CO phase, the proton PS moments are aligned uniformly; the net electric polarization emerges due to the proton configuration. 
Further increasing $g$, the first-order phase transition occurs from the FE CO phase to the AFE CO phase, where the alignments of the electric dipoles between the one-dimensional chains are changed from parallel to antiparallel configuration, and the macroscopic electric polarization disappears. 
\subsection{Charge and spin correlation functions}
In this section, we show the charge and spin correlations in the ground state beyond the mean-field approximation. 
We adopt the one-dimensional dimer-proton chain shown in Fig.~\ref{fig:lattice}(c) and analyze the Hamiltonian in Eq.~(\ref{eq:hamil}) by using the exact diagonalization method. 
Figures~\ref{fig:corr}(a)-\ref{fig:corr}(c) show the variations of the charge correlation function for the CDW state $N(k=1/2)$, that for the polar CO state $P(k=0)$, and the spin correlation function for the AFE DM state $S(k=1/2)$ in the $t_{\rm pro}$-$g$ planes. 
Schematic charge and spin configurations in these states are shown in Fig.~\ref{fig:1d}. 
We checked that the values of correlation functions at the other wave numbers are smaller than those shown in Fig.~\ref{fig:corr} in the whole parameter regions shown here. 

First, we focus on the results at $U=1$ shown in Fig.~\ref{fig:corr}(a).  
In the large-$g$ and small-$t_{\rm pro}$ region, the CDW correlation function $N(k=1/2)$ shows almost the maximum value $0.25$, indicating a realization of the CDW state shown in Fig.~\ref{fig:1d}(a). 
The AFM spin correlation function $S(k=1/2)$ is almost zero in this region, while it is remarkable in the small-$g$ and large $t_{\rm pro}$ region implying the AFM DM state. 
These results indicate that the CDW and AFM DM states are exclusive with each other. 
The polar CO correlation $P(k=0)$ is small, but shows a weak enhancement between the CDW and AFM DM states. 

At $U=2$ shown in Fig.~\ref{fig:corr}(b), the parameter region of the CDW state is shrunk, while that of the AFM DM state is extended. 
The polar CO correlation is enhanced in the small-$t_{\rm pro}$ and intermediate-$g$ region between the CDW and AFM DM states, in which the AFM correlation also shows the large value, implying the coexistence of the polar CO and AFM structures as shown in Fig.~\ref{fig:1d}(b). 
This means that the polar CO state is compatible with the AFM correlation in contrast to the CDW state. 
With further increasing $U$, the CDW correlation becomes smaller, while the polar CO correlation is dominant in the wide range of the large-$g$ region as shown in Fig.~\ref{fig:corr}(c). 
Although the AFM correlation is further enhanced in the whole parameter range, a weak suppression is seen in the region where the CDW correlation is enhanced. 

Through the results explained above, it is concluded that the CDW, polar CO, and DM states compete with each other in the ground state. 
The phase transitions between them are governed by the proton-electron coupling $g$, the proton tunneling $t_{\rm pro}$, and the local electron-electron interaction $U$. 
These results are consistent with the characteristic structure of the phase diagram in the three-dimensional lattice obtained by the mean-field approximation. 

We discuss the stabilities of these three states. 
For simplicity, we focus on the one-dimensional lattice and assume the isolated dimer limit given by $t=0$. 
First, we start with the case of the classical proton limit given by $t_{\rm pro} = 0$. 
We obtain the analytical expressions of the energies per unitcell (two dimers) in the three states as follows; 
\begin{align}
E^{\rm CDW} &= -\frac{1}{2}(g + 4t_{A}) + U_{\rm eff}, \label{eq:cdw}\\
E^{\rm CO} &= -\frac{1}{2}\sqrt{g^2 + 16t_{A}^2},  \label{eq:co}\\
E^{\rm DM} &= -2t_{A},
\end{align}
where $U_{\rm eff}$ is the intra-dimer Coulomb interaction between two holes in the antibonding orbital given by $U_{\rm eff} = 2t_{A}+(U+V_{A})/2-\sqrt{4t_{A}^2+(U-V_{A})^2/4}$. 
We find that the energy of the DM state is not lower than the others for finite value of $g$. 
In the noninteracting case ($U_{\rm eff} = 0$), the CDW state has the lower energy than the polar CO state for any finite values of $t_{A}$ and $g$. 
The stability of the CDW state can be understood by the molecular-orbital configurations as follows. 
Figure~\ref{fig:level} shows the energy levels of the molecular orbitals in the CDW state and the polar CO state.
At $t_{A} = 0$, the energy levels of the molecular orbitals represented by the bold broken lines in the figure are degenerate in the CDW and polar CO states, although the electron and proton configurations are different with each other. 
When $t_{A}$ is introduced, the molecular orbitals are hybridized and form the bonding and antibonding orbitals denoted by the bold solid lines in the figure. 
The lowering of the energy of the antibonding orbital in the CDW state is given by $t_A$, and this value is larger than that in the polar CO state given by $\sqrt{(g/4)^2+t_A^2}-g/4$ for any positive $t_A$. 
Therefore, the stability of the CDW state is attributed to the cooperation of the potential energy due to protons and the electronic kinetic energy. 
When we introduce the intra-dimer Coulomb interaction $U_{\rm eff}$, the polar CO state overcomes the CDW state because of $U_{\rm eff}$ in Eq.~(\ref{eq:cdw}). 
This is the reason why the polar CO state is dominant in the large-$U$ region in which the CDW state is suppressed. 
When we turn on the proton tunneling $t_{\rm pro}$, the $x$-component of the proton PS moment increases, while the $z$-component corresponding to the proton displacement reduces. 
Thus, the energy gains due to the proton-electron coupling in the CDW and polor CO states are suppressed. 
As a result, the DM state without the proton displacement is more stable than these states for large $t_{\rm pro}$. 
\section{Charge excitation}
\begin{figure*}[t]
\begin{center}
\includegraphics[width=2.0\columnwidth, clip]{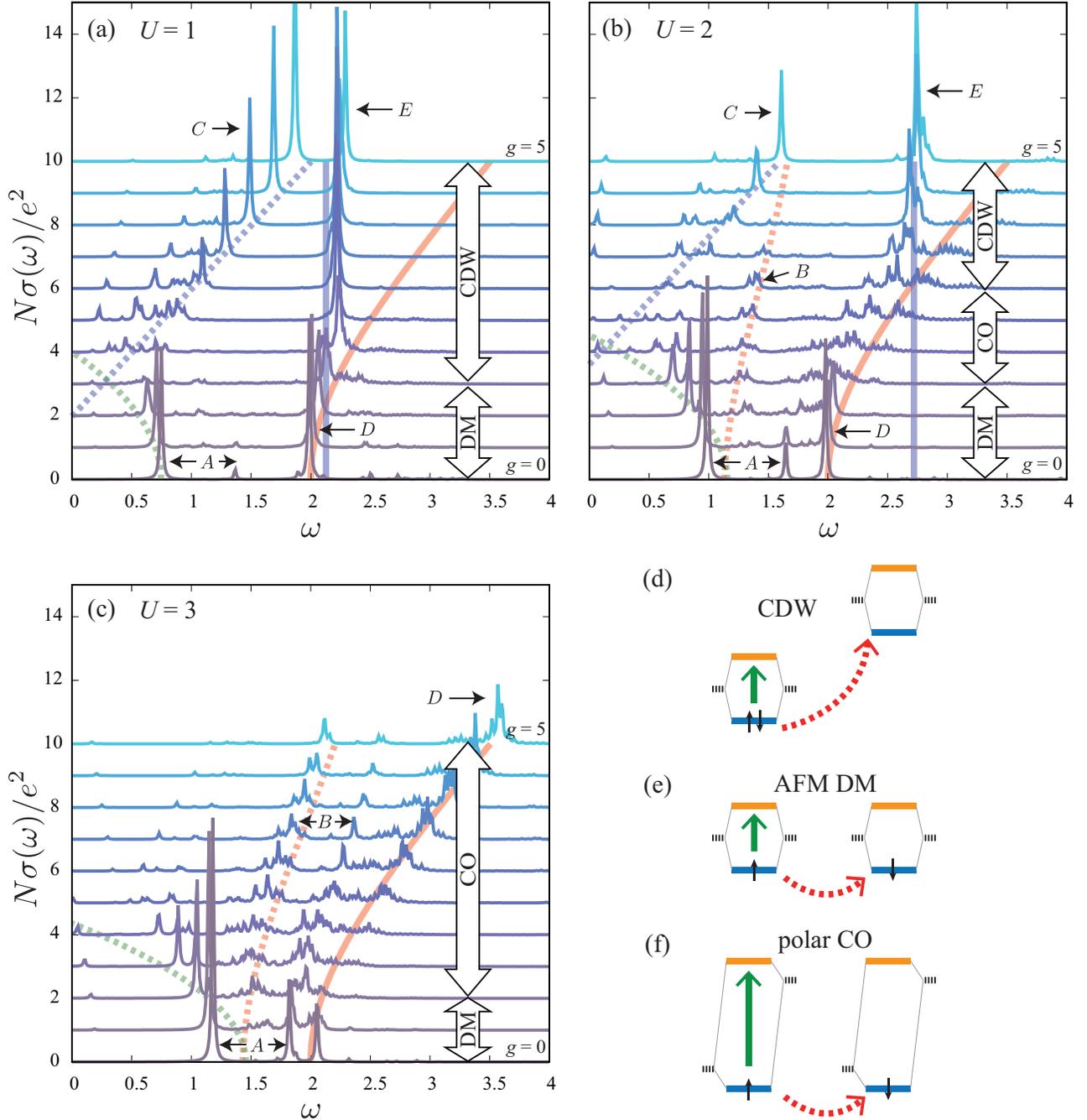}
\end{center}
\caption{
Optical conductivity spectra for several values of $g$ at (a) $U=1$, (b) $U=2$, and (c) $U=3$. We chose $t_{\rm pro}=0.1$. 
The numerical value of $g$ is changed from $0$ to $5$ by $0.5$ from the bottom to top. 
Solid and broken lines represent the excitation energies of the intra- and inter-dimer charge excitations given in (d)-(f) obtained in the isolated dimer limit [see text]. 
Schematic processes of the intra-dimer (solid arrow) and inter-dimer (broken arrow) charge excitations in (d) CDW, (e) AFM DM, and (f) CO states. 
}
\label{fig:sp}
\end{figure*}
%
%
%
In this section, we investigate the charge excitations in the one-dimensional lattice model. 
The proton tunneling amplitude is chosen as $t_{\rm pro}=0.1$. 
Figures~\ref{fig:sp}(a)-\ref{fig:sp}(c) show the optical conductivity spectra for several values of $g$ at $U=1$-$3$. 
We focus on the results at $U=2$, where the ground state is changed as AFM DM $\rightarrow$ polar CO $\rightarrow$ CDW with increasing $g$. 
Dominant five peaks are denoted by $A$, $B$, $C$, $D$, and $E$. 
First, the higher two peaks, $D$ and $E$, are focused on. 
With increasing $g$ from zero, the peak $D$ is harden, and split at around $g=3$. 
The higher energy peak is damped, and the lower peak is connected to the peak $E$. 
Above $g=3$, the energy of the peak $E$ does not show remarkable $g$ dependence. 
As for the lower energy peaks $A$, $B$, and $C$ in Fig.~\ref{fig:sp}(b), the peak $A$ is soften and dumped at around $g=1.5$, and the peak $B$ is harden with increasing $g$. 
In $g > 3$, the peak $C$ is remarkable, while the peak $B$ almost disappears. 

The optical spectra at $U=1$ is presented in Fig.~\ref{fig:sp}(a), where the CDW state is realized in $g > 1.5$ and the polar CO state does not appear. 
With increasing $g$, the amplitude of the peak $C$ is enhanced, and the energies of the peaks $D$ and $E$ weakly change. 
The spectra at $U=3$ are shown in Fig.~\ref{fig:sp}(c), where the polar CO state is realized in $g  > 1.5$ and the CDW state is absent. 
The peak $C$ and $E$ are not observed, while the peak $B$ and $D$ are remarkable.

We identify the excitations processes of the peaks $A$-$E$ as shown in Figs.~\ref{fig:sp}(d)-\ref{fig:sp}(f). 
Figure~\ref{fig:sp}(d) shows the excitation processes of the peak $C$ and $E$ observed in the CDW state. 
The peak $C$ is the charge transfer excitation in which a hole is transferred between the neighboring dimers. 
This process is denoted as ${\cal D}^2 {\cal D}^0$ $\rightarrow$ ${\cal D}^1 {\cal D}^1$, where ${\cal D}^m$ represents the lowest energy electronic state in a dimer with the hole number $m$. 
The peak $E$ is the local charge excitation from the antibonding to bonding orbitals described as ${\cal D}^2$ $\rightarrow$ ${\cal D}^{2*}$, where ${\cal D}^{m*}$ represents the excited state of ${\cal D}^{m}$. 
This is the so-called dimer excitation.~\cite{kezsmarki, faltermeier, hashimoto, naka_opt} 
Figures~\ref{fig:sp}(e) and \ref{fig:sp}(f) show the excitation processes in the AFM DM and polar CO states, respectively. 
The peaks $A$ and $B$ are identified as the inter-dimer charge excitations described by ${\cal D}^1 {\cal D}^1$ $\rightarrow$ ${\cal D}^0 {\cal D}^2$, which is the so-called Hubbard excitation,~\cite{kezsmarki, faltermeier, hashimoto, naka_opt} and the peak $D$ is assigned as the dimer excitation described by ${\cal D}^1$ $\rightarrow$ ${\cal D}^{1*}$. 

We evaluate the excitation energies in an isolated dimer, i.e. $t=0$, and analyze the $g$ and $U$ dependences of the optical spectra. 
The results are shown in Figs~\ref{fig:sp}(a)-\ref{fig:sp}(c). 
For simplicity, we set  $t_{\rm pro}=0$ and $P^{z} = 1/2$ in the polar CO and CDW states, assuming that the hydrogen bonds are fully polarized. 
The inter-dimer Coulomb interaction $V$ is treated as the perturbation up to the order of ${\cal O}(V)$.  

First, we focus on the dimer excitations. 
The energy of the dimer excitation in the polar CO state shown in Fig.~\ref{fig:sp}(f) is given by 
\begin{align}
\Delta_{\rm dimer}^{\rm CO} &= \frac{1}{2}\sqrt{g^2+16t_{A}^2} + \frac{g^2V}{g^2+16t_{A}^2}, \label{eq:delD}
\end{align}
in the case of $t=0$ and $t_{\rm pro}=0$. 
The first term originates from the energy difference between the antibonding and bonding orbitals, and the second term is due to the energy cost caused by the electric dipole-moment reversal inside the dimer. 
The energy of the dimer excitation in the CDW state shown in Fig.~\ref{fig:sp}(d) is given by 
\begin{align}
\Delta_{\rm dimer}^{\rm CDW} &= U - U_{\rm eff} + 2t_{A}. \label{eq:delE}
\end{align}
The first term corresponds to the energy of the spin singlet final state where each of the antibonding and bonding orbitals is singly occupied, and the sum of the second and third terms corresponds to the energy of the initial state where the antibonding orbital is doubly occupied. 
It is shown that $\Delta_{\rm dimer}^{\rm CO}$ is proportional to $g^{2}$ for small $g$, and coincides with the excitation energy from the bonding to antibonding orbital inside a dimer in the AFM DM state. 
On the other hand, $\Delta_{\rm dimer}^{\rm CDW}$ does not depend on $g$. 
The energies of the peaks $D$ and $E$ obtained by the numerical calculation are well reproduced by the above expressions of $\Delta_{\rm dimer}^{\rm CO}$ and $\Delta_{\rm dimer}^{\rm CDW}$, respectively, as shown in Figs.~\ref{fig:sp}(a)-\ref{fig:sp}(c). 

Next, the Hubbard excitation in the AFM DM and CO states and the charge transfer excitation in the CDW state are focused on. 
The energy of the Hubbard excitation in the AFM DM state is obtained by the perturbation with respect to $g$ up to the order of ${\cal O}(g^2)$. 
The result is given by 
\begin{align}
\Delta_{\rm Hubb}^{\rm DM} = U_{\rm eff} - \gamma g^2 + \frac{1}{4} V, \label{eq:delA}
\end{align}
where the first term corresponds to the Hubbard excitation energy in the AFM DM state at $g=0$, and the second and third terms are correction by $g$ and $V$, respectively. 
A positive constant $\gamma$ is given by Eq.~(\ref{eq:Hubb_DM}) in Appendix, and is given as a function of $t_A$, $U$, and $V_A$. 
The energy of the Hubbard excitation in the polar CO state is obtained as an inter-dimer excitation energy between the antibonding orbitals under the charge disproportionation inside of dimers. 
The explicit form is given in Eq.~(\ref{eq:Hubb_CO}) in Appendix. 
Finally, the energy of the charge transfer excitation in the CDW state is given by 
\begin{align}
\Delta_{\rm CT}^{\rm CDW} = \frac{1}{2}g - U_{\rm eff} + \frac{3}{4}V, 
\end{align}
where the first term represents the energy difference of the antibonding orbitals in the neighboring dimers under the proton displacements, the second term is the effective intra-dimer Coulomb interaction between the two holes in the same antibonding orbital, and the third term is due to the inter-dimer Coulomb interaction. 

As shown in the expressions of $\Delta_{\rm Hubb}^{\rm DM}$ and $\Delta_{\rm CT}^{\rm CDW}$, the Hubbard excitation in the DM state and the charge transfer excitation in the CDW state are soften and harden, respectively, with increasing $g$. 
It is worth to note that the softening of the Hubbard excitation is due to the screening of the intra-dimer Coulomb interaction $U_{\rm eff}$ by the quantum proton motion., that is the so-called bi-polaron effect.~\cite{alexandrov} 
The analytical results of $\Delta_{\rm Hubb}^{\rm DM}$ and $\Delta_{\rm CT}^{\rm CDW}$ denoted by the green and blue broken lines in Figs.~\ref{fig:sp}(a)-\ref{fig:sp}(c), respectively, well explain the numerical calculation results of the peaks $A$ and $C$. 
The Hubbard excitation in the polar CO state ($\Delta_{\rm Hubb}^{\rm CO}$) denoted by the red broken line in Figs.~\ref{fig:sp}(b) and \ref{fig:sp}(c) is harden slightly with increasing $g$. 
This is attributable to the increase of $U_{\rm eff}$ due to the increase of the intra-dimer charge disproportionation of the antibonding orbital. 
The energy of the peak $B$ obtained numerically is well reproduced by $\Delta_{\rm Hubb}^{\rm CO}$. 

\section{Discussion}
We compare the present numerical calculations with the experimental observations in $\kappa$-{\it X}$_3$(Cat)$_2$.
First, we discuss the isotope effect of proton. 
As mentioned in Sec. I, by the substitution of H $\rightarrow$ D, the low-temperature phase is changed from the DM to the CDW phase accompanied by localization of the deuterons. 
The isotope effects have been studied so far in the hydrogen-bonded ferroelectrics, and the following two effects have been proposed. 
1) Elongation of the hydrogen bond length, as well as the increase of the mass of {\it X} ion, reduces the quantum tunneling.~\cite{brinc, degennes} 
2) Increasing of the lattice deformation surrounding the {\it X} ion is caused by strengthening of the coupling between the lattice and {\it X}.~\cite{yamada} 
In the present model Hamiltonian, 1) and 2) introduced above correspond to decrease of the proton tunneling $t_{\rm pro}$ and increase of the proton-electron coupling $g$, respectively. 
Here, we assume that the energy of the molecular orbitals are affected by the lattice deformations. 
Both of the two kinds of the parameter changes promote the phase transition from the DM to CDW phase. 
This tendency is consistent with the experimental results.~\cite{ueda} 

Next, we discuss the correspondence between the several phases in the calculation and the experimental observations. 
Through the first-principles calculation and the analysis of the dielectric constant measured in $\kappa$-H$_3$(Cat)$_2$, the magnitude of $t_{\rm pro}$ is estimated as about $0.01 t_{A}$-$0.1 t_{A}$.~\cite{shimozawa} 
It is known that in the several DM systems, the intra-molecular Coulomb interaction $U$ is at about $2 t_A$-$3 t_A$.~\cite{fukuyama, kino, kanoda} 
From the calculated phase diagram shown in Figs.~\ref{fig:pd}(b) and \ref{fig:pd}(c), we estimate $t_{\rm pro}$ in $\kappa$-X$_3$(Cat)$_2$ to be around $0.1 t_A$, since the DM and CDW phases compete with each other in this parameter region, in the similar way to the experimental observation. 
The absence of the AFM long range order in $\kappa$-H$_3$(Cat)$_2$ will be discussed later. 
We note that the polar CO phase predicted by the calculation has not been observed experimentally yet. 
This discrepancy might be attributable to the structural change in $\kappa$-D$_3$(Cat)$_2$ as follows. 
According to the extended Huckel calculations, the intra-dimer electron transfer $t_{A}$ below $T_{\rm CDW}$ is about $50$ \% larger than that in the high-temperature DM phase.~\cite{ueda_band} 
This is due to the reduction of the distance between the two molecules in the hole-rich dimer. 
This enhancement of the transfer integral increases the stability of the CDW phase in $\kappa$-D$_3$(Cat)$_2$, in comparison with the polar CO phase as discussed in Sec. IV. 
We expect that the polar CO phase is realized by applying the magnetic field, since the local spin moments survive in this phase, in contrast to the CDW phase. 
In the polar CO phase, novel magnetoelectric phenomena are expected due to the intra-dimer multipole composed of the magnetic and electric dipole moments.~\cite{naka_me} 

We discuss a possible scenario of the quantum spin liquid state observed in $\kappa$-H$_3$(Cat)$_2$. 
Theoretical realizations of the spin liquid state are beyond the present mean-field approximation and the small-cluster calculations. 
However, it is found that the CDW fluctuation suppresses the AFM spin correlations as shown in Figs.~\ref{fig:corr}(a)-\ref{fig:corr}(c). 
We suppose that the charge degree of freedom plays essential roles on the realization of the spin liquid state in $\kappa$-H$_3$(Cat)$_2$. 
Experimentally, an increase of the dielectric constant with decreasing temperature, which is considered as a precursor of the CDW state, is observed in low temperatures below of $\kappa$-H$_3$(Cat)$_2$.~\cite{shimozawa} 
Recently, it is found that the electronic and structural phases in low temperatures below $50$ K in $\kappa$-H$_3$(Cat)$_2$ strongly depends on the samples. 
The DM phase without the long-range magnetic order remains until $50$ mK in some samples, and the CDW phase is realized through the first-order phase transition at around $50$ K in others. 
These facts suggest that $\kappa$-H$_3$(Cat)$_2$ is located in the DM phase near the boundary of the CDW phase, and the large CDW fluctuation which suppresses the AFM order is expected. 
Roles of the electronic charge fluctuation on realization of the spin liquid state have been also stressed in 
other spin-liquid candidates of the dimer Mott insulators, e.g.  
$\kappa$-(BEDT-TTF)$_2$Cu$_2$(CN)$_3$ and EtMe$_3$Sb[Pd(dmit)$_2$]$_2$.~\cite{shimizu, itou} 
In the present Hydrogen-based molecular conductor system, the quantum proton motion reduces the effective electron-electron repulsions in the molecular dimers and promotes the electron-proton entangled charge fluctuation. 
As shown in Figs.~\ref{fig:sp}(a)-\ref{fig:sp}(c), this kind of charge excitations can be examined directly by the optical spectra. 

Finally, we propose an experimental method to evaluate the amplitude of the proton-electron coupling constant $g$. 
By using Eq.~(\ref{eq:delD}), the proton-electron coupling is given by $g \sim 2 \sqrt{E_{\rm dimer}^{\rm CO} - 4 t_A^2}$. 
Here, the second term of Eq.~(\ref{eq:delD}) is omitted, because this is sufficiently smaller than the first term for the present parameter set. 
Since the intra-dimer transfer integral can be estimated by the first-principles calculation,~\cite{tsumuraya} 
the proton-electron coupling amplitude is evaluated through the optical measurement of the dimer excitation. 
The comparison of the values of $g$ in $\kappa$-H$_3$(Cat)$_2$ and $\kappa$-D$_3$(Cat)$_2$ enables us to elucidate whether the driving force of the CDW transition is the change of $g$ or $t_{\rm pro}$. 

\section{Summary}
In summary, we have presented a microscopic theory of properties of electronic and protonic states in the hydrogen-bonded molecular conductors, motivated by the recent experimental studies in $\kappa$-{\it X}$_3$(Cat)$_2$. 
We have introduced an effective model for $\kappa$-{\it X}$_3$(Cat)$_2$ and obtained the ground-state phase diagram and charge excitation spectra. 
The three competing electronic and protonic phases appear: the CDW, polar CO, AFM DM phases. 
There are mainly two optical excitation modes in the DM phase, i.e. the dimer and Hubbard excitations, which are harden and soften, respectively, with increasing the proton-electron coupling constant $g$. 
The softening of the Hubbard excitation implies reduction of ``Hubbard U'' due to the quantum proton motion.
This result provides us a direct experimental method to evaluate the magnitude of the proton-electron coupling. 
The present theory for $\kappa$-{\it X}$_3$(Cat)$_2$ triggers further progresses of the microscopic comprehensions of other proton-electron coupled materials. 

The authors would like to thank K. Hashimoto, H. Mori, T. Sasaki, H. Seo, and M. Shimozawa for valuable discussions. 
This work is supported by Grants-in-Aid for Scientific Research (No. 15H02100, 16K17731 and 17H02916) from MEXT (Japan). 

\section{Appendix}
In this Appendix, we present details of the analytical formulas of the Hubbard excitation energies in the DM and polar CO states. 
We assume, for simplicity, $t=0$ in the one-dimensional lattice model. 

First, the Hubbard excitation in the DM state is examined. 
We adopt the perturbational approximation with respect to the proton-electron coupling $g$ and the inter-dimer Coulomb interaction $V$ up to the orders of ${\cal O}(g^2)$ and ${\cal O}(V)$. 
The initial and final states of the excitations are expressed by $\cdots {\cal D}^1{\cal D}^1{\cal D}^1{\cal D}^1 \cdots$ and $\cdots {\cal D}^1{\cal D}^0{\cal D}^2{\cal D}^1 \cdots$, respectively. 
The protons are assumed to occupy the antibonding states in the hydrogen bonds. 
The wave function for the unperturbed initial state is given by 
\begin{align}
| i \rangle = \prod_{i=1}^N | \alpha_{i \sigma} \rangle | -_i^x \rangle. 
\label{eq:initial}
\end{align}
We introduce that $| \alpha(\beta)_{i \sigma} \rangle$ and $| -(+)_i^x \rangle$ are the antibonding (bonding) states of the hole in the $i$-th dimer and that of the proton in the $i$-th hydrogen bond, respectively. 
We set that an excitation of a hole occurs between the $i=1$ and $i=2$ dimers. 
The wave function for the unperturbed final state is given by 
\begin{align}
| f \rangle = | 0_1 \rangle | d_2^- \rangle \prod_{i=3}^N | \alpha_{i \sigma} \rangle \prod_{i=1}^N | -_i^x \rangle, 
\end{align}
where $| 0 \rangle$ and $| d^- \rangle$ are the unoccupied and doubly occupied states by holes, respectively. 
We define  
\begin{align}
| d^\pm \rangle = C^\pm_{\alpha} | \alpha_\uparrow \alpha_\downarrow \rangle + C^\pm_{\beta}| \beta_\uparrow \beta_\downarrow \rangle. 
\end{align}
The coefficients are given by 
\begin{align}
C^\pm_{\alpha} &= - \frac{U+2t_A-\lambda^\pm}{\sqrt{2} L^\pm}, \\
C^\pm_{\beta} &= \frac{U-2t_A-\lambda^\pm}{\sqrt{2} L^\pm}, 
\end{align}
where $L^\pm=\sqrt{4t_A^2+(\lambda^\pm - U)^2}$. 
We introduce the eigenvalue of $\left| d^{\pm} \right\rangle$ as $\lambda^{\pm} = (U+V_{\rm A})/2 \pm \sqrt{4t_{A}^2 + (U-V_{\rm A})^2/4}$. 

The perturbational terms of the Hamiltonian are given by 
\begin{align}
{\cal H}_g = \frac{1}{2} g \sum_{i=1}^N (n_{i+1 a} - n_{i b})P_i^{z},  
\end{align}
and 
\begin{align}
{\cal H}_V = V \sum_{i=1}^N n_{i+1 a} n_{i b}. 
\label{eq:hv}
\end{align}
The lowest order perturbation energy is given by 
\begin{align}
\widetilde{E}_{\mu} = E_{\mu} + \sum_{\nu(\neq \mu)} \frac{|\langle \nu| {\cal H}_g | \mu \rangle|^2}{E_{\mu} - E_{\nu}} + \langle \mu| {\cal H}_V | \mu \rangle, 
\end{align}
where $| \mu \rangle$ and $| \nu \rangle$ represent the initial and intermediate states, and $E_{\mu}$($E_{\nu}$) is the unperturbed energy of $| \mu \rangle$($| \nu \rangle$). 
The energies of the orders of ${\cal O}(g)$ and ${\cal O}(gV)$ are zero. 
Then, we have the Hubbard excitation energy $\Delta_{\rm Hubb}^{\rm DM}$($\equiv \widetilde{E}_{f} - \widetilde{E}_{i}$) as 
\begin{align}
\Delta_{\rm Hubb}^{\rm DM} 
&= U_{\rm eff} + \frac{1}{4} V - g^2 \Biggl[ \frac{1}{16 t_{\rm pro} + 8(U - \lambda^-)} \frac{4t_{\rm A}^2}{(L^-)^2} \notag \\
&+ \frac{1}{16 t_{\rm pro} + 8(\lambda^+ - \lambda^-)} \left( \frac{L^+}{L^-} \right)^2 \frac{4(U-\lambda^-)^2}{(\lambda^+ - \lambda^-)^2} \notag \\
& - \frac{1}{32(t_{\rm A} + t_{\rm pro})} + \frac{1}{64t_{\rm pro}} + \frac{1}{16t_{\rm pro}} \frac{(U-V_{\rm A})^2}{(\lambda^+ - \lambda^-)^2} \Biggr], 
\label{eq:Hubb_DM}
\end{align}
where the coefficient of $g^2$ in the third term corresponds $\gamma$ in Eq.~(\ref{eq:delA}) in Sec. V. 

Next, the Hubbard excitation energy in the polar CO state is examined. 
We assume that the protons are fully polarized in the hydrogen bonds, and set $t_{\rm pro}=0$ 
The inter-dimer Coulomb interaction in Eq.~(\ref{eq:hv}) is treated as the perturbational term of the Hamiltonian. 
The unperturbed Hamiltonian for single dimer is given by 
\begin{align}
{\cal H}_{\rm 0} 
&= \frac{1}{2}g (n_{b} - n_{a}) + t_{A}\sum_{\sigma} \left( c_{a \sigma}^\dagger c_{b \sigma} + {\rm H.c.} \right) \notag \\
&+ U \sum_{\mu} n_{\mu \uparrow} n_{\mu \downarrow} + V_{A} n_{a} n_{b}, 
\label{eq:h0}
\end{align}
where the first term represents the electrostatic potential originating from the protons. 
The unperturbed initial and final states are given by 
\begin{align}
| i \rangle = \prod_{i=1}^N | {\hat \alpha}_{i \sigma} \rangle, 
\end{align}
and
\begin{align}
| f \rangle = | 0_1 \rangle | {\hat d}_2^- \rangle \prod_{i=3}^N | {\hat \alpha}_{i \sigma} \rangle, 
\end{align}
respectively. 
We introduce $| {\hat \alpha}_{\sigma} \rangle$ and $| {\hat d}^- \rangle$ as the lowest energy states for Eq.~(\ref{eq:h0}), where the numbers of holes in the dimer are one and two, respectively. 
The perturbation energy is given by 
\begin{align}
\widetilde{E}_{\mu} = E_{\mu} + \langle \mu| {\cal H}_V | \mu \rangle, 
\end{align}
where $E_{\mu}$ and $| \mu \rangle$ represent the unperturbed energy and wave function, respectively. 
The Hubbard excitation energy is obtained as 
\begin{align}
\Delta_{\rm Hubb}^{\rm CO} = {\widetilde E}_{f} - {\widetilde E}_{i}. 
\label{eq:Hubb_CO}
\end{align}


\end{document}